\documentclass[twocolumn,aps,prb,amsmath]{revtex4}

\usepackage{graphicx}
\usepackage{epsfig}
\usepackage{amsmath}
\usepackage{amsfonts}
\usepackage{amssymb}

\def\be{\begin{equation}}
\def\ee{\end{equation}}
\def\bfi{\begin{figure}}
\def\efi{\end{figure}}
\def\bea{\begin{eqnarray}}
\def\eea{\end{eqnarray}}
\begin{document}
\title{Reentrant phase diagram and pH effects in
cross-linked gelatin gels}
\author{T. Abete$^{(a)}$, E. Del Gado$^{(b)}$,
L. de Arcangelis$^{(c)}$\footnote{to whom
correspondence should be addressed: dearcangelis@na.infn.it.},
D. Hellio Serughetti$^{(d)}$\footnote{present address: Saint Gobain Recherche,
39 quai Lucien Lefranc, 93303 Aubervilliers Cedex, France},
M. Djabourov$^{(d)}$} \affiliation{ $^{(a)}$
Department of Physical Sciences and CNISM,
University of Naples ``Federico II'', 80125 Napoli, Italy\\
$^{(b)}$
Institute for Polymers, ETH, 8093 Z\"urich, Switzerland\\
$^{(c)}$
Department of Information Engineering and CNISM,
Second University of Naples, 81031 Aversa (CE), Italy\\
$^{(d)}$
Laboratoire de Physique Thermique, ESPCI, 75231 Paris Cedex 5, France
}

\begin{abstract}
Experimental results have shown that the kinetics of bond formation
in chemical crosslinking of gelatin solutions are strongly affected
not only by gelatin and reactant concentrations but also by the
solution pH. We present an extended numerical investigation of the
phase diagram and of the kinetics of bond formation as a function of
the pH, via Monte Carlo simulations of a lattice model for gelatin
chains and reactant agent in solution. We find a reentrant phase
diagram, namely gelation can be hindered either by loop formation at
low reactant concentrations, or by saturation of chain active sites
via formation of single bonds with a crosslinker at high reactant
concentrations. The ratio of the characteristic times for the
formation of the first and the second bond between the crosslinker
and an active site of a chain is found to depend on the reactant
reactivity, in good agreement with experimental data.
\end{abstract}

\maketitle Gelatin gels have received great attention \cite{gelatin}
because of the interesting fundamental aspects of their rheological
behavior and their numerous industrial applications. When a
semi-diluted gelatin solution is cooled below room temperature,
chains start to form triple helices and progressively a connected
network is built. By adding crosslinker molecules to the gelatin
solution above the temperature of the coil-helix transition, the
thermoreversible gelation of helices can be avoided and a chemical
gel network is formed, as chain aminoacids react with crosslinkers.
This possibility of chemical gelation in gelatin solutions is
extremely relevant for the numerous applications in pharmaceutical,
photographic and biomedical industries. The physical gel is
characterized by an extreme biodiversity due to chemical composition
of the native collagen, molecular weight distribution, solution
properties such as concentration or pH, which may affect the
temperature of helix formation \cite{mad1}. In addition, the option
of chemical gelation or a combination of chemical and physical
gelation represent an extremely interesting way of enhancing the
performance of these materials, in terms of structural and
rheological features. As a consequence, an important issue for the
development of technological applications is the understanding and
the governance of the interplay of the different control parameters
(gelatin concentration, crosslinker concentration, pH or
temperature).

Recently, extended studies have been performed on gelatin in
solution with bisvinylsulphonemethyl (BVSM) reactant \cite{mad3},
able to establish bifunctional covalent bonds with the lysine, the
hydroxylysine and possibly with other amine groups of gelatin
chains. The influence on bond formation of various parameters, i.e.
the concentration of gelatin and reagent as well as the solution pH,
was investigated. Microcalorimetry measurements were able to monitor
the development of the chemical reaction in time by detecting the
exothermic enthalpy change during the formation of $C-N$ bonds. The
kinetics of cross-link formation was found to follow a double
exponential decay with two characteristic times. However, when
counting the number of cross-links binding BVSM and gelatin, the
method could not discriminate between bonds established by free
reactants with a chain, bonds leading to a bridge between two
gelatin chains, or bonds leading to a loop within a chain. This lack
of information on the kinetics clearly hindered the characterization
of the structure and therefore the understanding of the mechanical
properties of the gel, requiring an alternative approach to reach a
deeper comprehension. A joint investigation, pursued by
complementing the experimental study with suitably designed
numerical simulations, allowed to shed some light on these
unanswered questions.

By means of Monte Carlo simulations on the cubic lattice of a simple
model we analyzed the kinetics of bond formation in chemical
cross-linking of a gelatin solution \cite{noi}. We considered a
solution of polymer chains at different concentrations where
reactant monomers can diffuse and form bonds with the active sites
along the chains, producing the cross-linking. It was possible to
follow the kinetics of the gel formation by varying the gelatin
concentration, the cross-linker concentration and its bonding
probability (i.e. reactivity). The numerical data were able to
reproduce extremely well the experimental findings. The combined
analysis showed that the two time scales detected in the experiments
correspond to the average times for forming, respectively, single
bonds reactant-chain and bridges chain-chain via cross-linkers. We
related these two times to the characteristic times of diffusion of
free reactants and reactants which have already formed one bond with
a chain. Their ratio controls the kinetics of bond formation:
varying the concentration, the cross-linker reactivity and the pH
strongly affects this ratio and therefore the kinetics of the
gelation process. We could also show that the reaction rate for a
reactant to form a bridge between two active sites allows to finely
tune the kinetics of gelation by affecting the ratio of the two
characteristic times. This can be achieved in experiments, for
instance, by changing the reactant agent in the gelatin solution.

Another crucial question put forward by the experimental study
\cite{mad3} is the role of the pH. In particular, a qualitative
change of the gelation kinetics was observed by changing the pH,
which could not be explained, also due to the limited access to
structural information in the experiments: The ratio of the two
characteristic times experimentally appears to increase with the
solution pH. In fact, a higher pH activates more amine groups able
to react with BSVM along the gelatin chains. This can be mimicked in
the simulation by varying the number of active sites along the
chains. Preliminary data obtained in simulations \cite{noi} showed
some inconsistencies with the experimental observations for gelatin
solutions with BSVM. Indeed, the possibility of detecting different
regimes and varying meaningfully the number of active sites along
the chains strongly relies on the use of long enough chains. In this
paper we present therefore a new detailed numerical study of the
kinetics of the bond formation as a function of the pH and of the
concentration of gelatin and reactant. Here longer chains and larger
system sizes are considered, with the specific aim of modeling
solutions with different pH and exploring more extended regions of
the phase diagram, with respect to the previous study. We are able
to show that the gelation line exhibits a non-monotonic behavior,
that is, for high reactant concentrations the system goes back into
the sol phase for all pH values. This result is of course extremely
relevant for applications and is indeed in agreement with
experimental findings, where gelation is not observed even at
relatively high cross-linker concentrations \cite{lei,shi}. We show
that this feature is due to an enhanced probability of saturating
all the active sites by a single bond with a different crosslinker.
In specific conditions of high crosslinker concentrations, this can
actually hinder gelation, resulting in a reentrant phase diagram. By
systematically studying the dependence of the ratio of the two times
on the gelatin concentration for different pH, we are able to
reproduce and rationalize the experimental data. These indeed
correspond to a range of cross-linker reactivity lower than the one
investigated in the previous study. The present results lead to a
coherent scenario for the gelation kinetics, which exhaustively
complements the experimental study.

The paper is organized as follows: In section \ref{num} the
numerical model and the numerical simulations are described in
detail. The results on the gelation phase diagram are discussed in
section \ref{pd}, where also the number of loops and single bonds
formed at the end of the reaction is analyzed, together with the
size of the macromolecule. This last feature is not accessible
experimentally and is clearly extremely relevant for the mechanical
properties of the gel. In section \ref{bond} we study the kinetics
via the ratio of the two characteristic times as a function of
gelatin and reactant concentrations, as well as of the solution pH
and reactant reactivity. Concluding remarks are given in section
\ref{conclu}.

\section{Model and numerical study}
\label{num} We perform Monte Carlo simulations on a cubic lattice of
a system made of bi-functional monomers. Each monomer represents a
unit on the chain or else a reactant. The chains are formed by a
sequence of $n=20$ linked monomers. One monomer of the chain models
a Kuhn segment \cite{Doi}, and therefore represents more units. The
length of a Kuhn segment in a gelatin chain has been measured
\cite{Kuhn} to be of the order of $40$ {\AA}, corresponding to about
$10$ amino-acids. As compared to the experiments, our chains
correspond to shorter gelatin chains, containing only about $200$
amino-acids. Each monomer occupies simultaneously the eight sites of
the lattice elementary cell and, to take into account excluded
volume interactions, two occupied cells cannot have any site in
common. A fixed number of monomers along the chain are active sites
which may bind to the reactant in order to form complex clusters of
chains leading to the formation of a gel. The active sites are
tetra-functional: Two bonds are formed with the neighbors along the
chain and two are not saturated at the beginning of the simulation.
The number of active sites per chain, $n_{as}$, corresponds to a
fixed pH of the solution. Although the number of amine groups in a
gelatin chain actually linked to reactant cannot be measured
experimentally, it is estimated that at most a fraction of $20\%$
can react. Therefore we perform simulations for $n_{as}$ varying
from 5 to 11, which corresponds to a fraction of about $5\%$ to
$10\%$ active amino-acids in our chain.

Chains are randomly distributed on the lattice and diffuse via
random local movements using the bond-fluctuation dynamics
\cite{BFD,noi}. After equilibration, we add the reactant to the
system and let the solution diffuse towards the stationary state.
Due to the diffusion of cross-linkers and chains, when a free
reactant finds a nearest neighbor unsaturated active site, the first
bond forms along lattice directions. The second bond is instead
formed with probability $p_b\leq1$, since its formation might
require to overcome a free energy barrier \cite{noi,malla},
depending on the nature of the solution, the active sites and the
reactant. In particular, in these systems the rigidity of the $C-N$
link may significantly limit the angle between two bonds or the
effective chemical reactivity of the cross-linker may undergo
meaningful variation once the first bond is formed. The process goes
on until all the possible bonds are formed.

We perform numerical simulations of the model for lattice size
$L=200$, where the unit length is the lattice spacing $a=1$, with
periodic boundary conditions. The chain concentration $C$ and the
cross-linker concentration $C_r$ are defined as the ratio between
the number of present monomers and the maximum number of monomers
$N_{max}=L^3/8$ in the system. Using the percolation approach we
identify the gel phase as the state in which there is a percolating
cluster \cite{Flory_perco}. For a fixed set of parameters we
generate a number of configurations of the system and monitor the
reaction. In order to locate the gelation transition we analyze the
percolation probability $\Pi$, defined as the fraction of
configurations leading to a percolating cluster, and we identify the
transition with the line $\Pi=0.5$ \cite{stauffer}.

We investigate the behavior of the number of bonds formed during the
reaction process and we classify the bonds in three
categories: \\
1. Bonds between a free reactant and an
active site (we will refer to this type of bond as
\emph{single-bonds}); \\
2. Bonds between a linked reactant and an active site
of another chain (\emph{bridges});\\
3. Bonds between a reactant and two active sites of the same chain
(which in the following we will call \emph{loops}).
\\
We analyze the kinetics of bond formation varying the chain
concentration $C$, the cross-linker concentration $C_r$, the
probability $p_b$ of second bond formation and the number of active
sites $n_{as}$. The time is measured in Monte Carlo unit time, i.e.
the attempt to move all monomers once. The number of bonds is
normalized by the maximum number of bonds that can be formed
depending on the limiting agent of the simulation: The normalization
factor is the minimum value between twice the number of reactant
monomers and twice the number of active sites in the system.

\section{Phase diagram and loops}
\label{pd} 
\begin{figure}
\includegraphics[width=8cm]{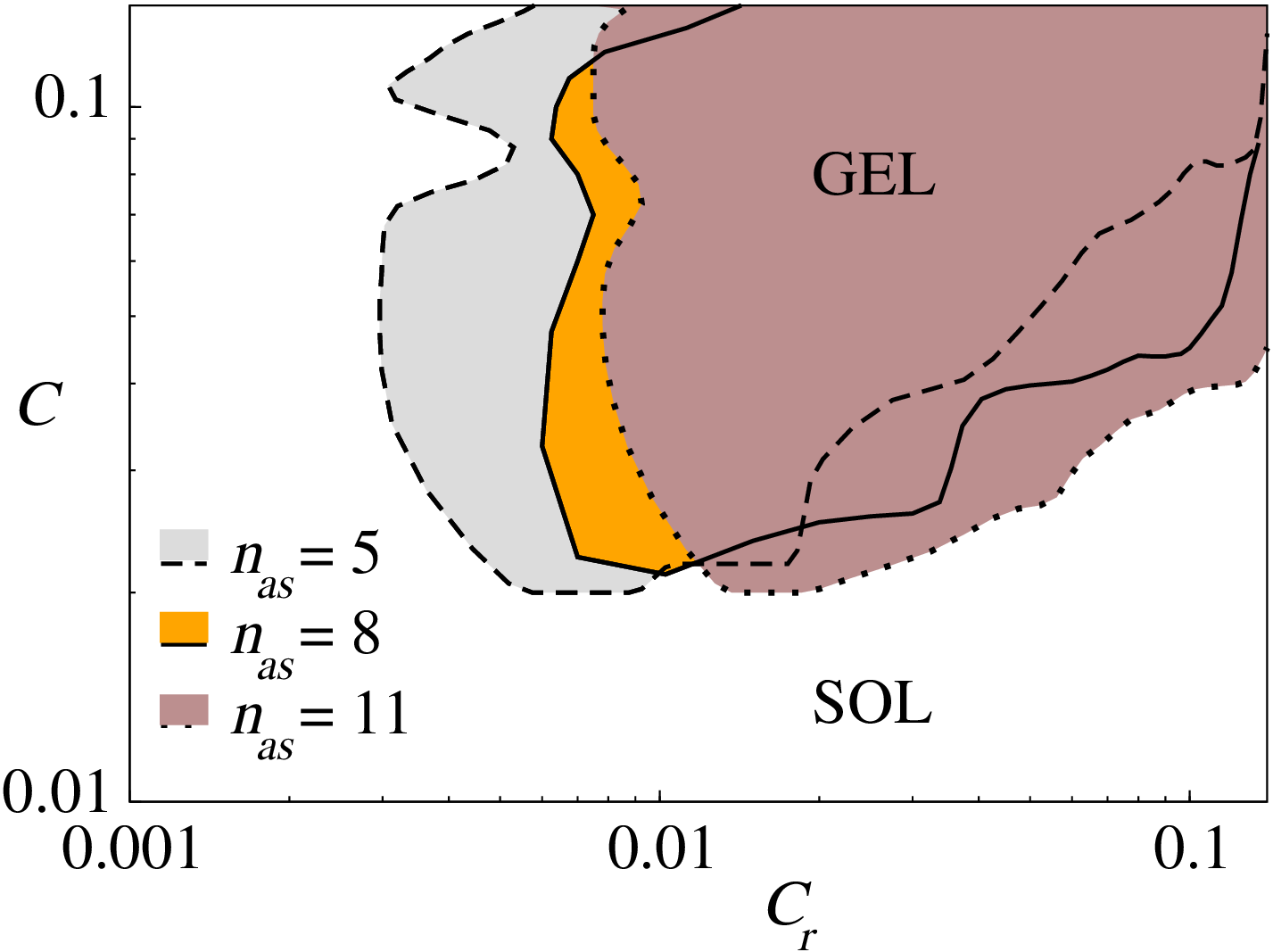}
\caption{ The phase diagram, obtained plotting the percolation
probability $\Pi$ as a function of chain and reactant concentration,
$C$ and $C_r$ respectively, for $C$ ranging from $C=0.01$ to
$C=0.14$ and $C_r$ ranging from $C_r=0.001$ to $C_r=0.14$. The
spanning probability has been averaged over 32 independent initial
configurations of a sample of size $L=200$ for $n_{as}=5,8,11$ and
$p_b=1$.} \label{fig1}
\end{figure}

We first determine a qualitative phase diagram by varying
the chain and cross-linker concentrations, $C$ and $C_r$
respectively, for $n_{as}=5,8,11$, which correspond to different
values of the pH. The results are shown in Fig.\ref{fig1}.

Following the standard site-bond percolation picture \cite{stauad},
one expects that the gel phase is obtained provided that enough
crosslinkers are added. However, it has been observed in experiments
that this is not always the case, and that this feature is affected
by the value of the pH. Fig.\ref{fig1} shows that for all values of
$n_{as}$ gelation can be obtained in a limited range of $C_r$.
Indeed, the figure clearly indicates that, at a given gelatin
concentration $C$, increasing $C_{r}$ beyond a certain value
corresponds to a strong decrease of percolation probability and
hinders the presence of a gel at the end of the reaction. Of course
this {\it reentrant} phase diagram is the product of the balance
among gelatin concentration, crosslinker concentration and number of
active sites per chain. Our reentrant phase diagram is in agreement
with experimental findings for polymer aqueous solutions \cite{lei,
shi}, where gelation can be inhibited by different mechanisms,
namely polyelectrolyte effects due to the complexation process.

If the number of active sites per chain increases, $n_{as}=5, 8,
11$, i.e. the pH increases,
the gel phase moves towards regions with higher concentration of
cross-linkers.
This suggests that in all cases, for sufficiently
high $C_{r}$, the system goes back into the sol phase.

\begin{figure}
\includegraphics[width=7cm]{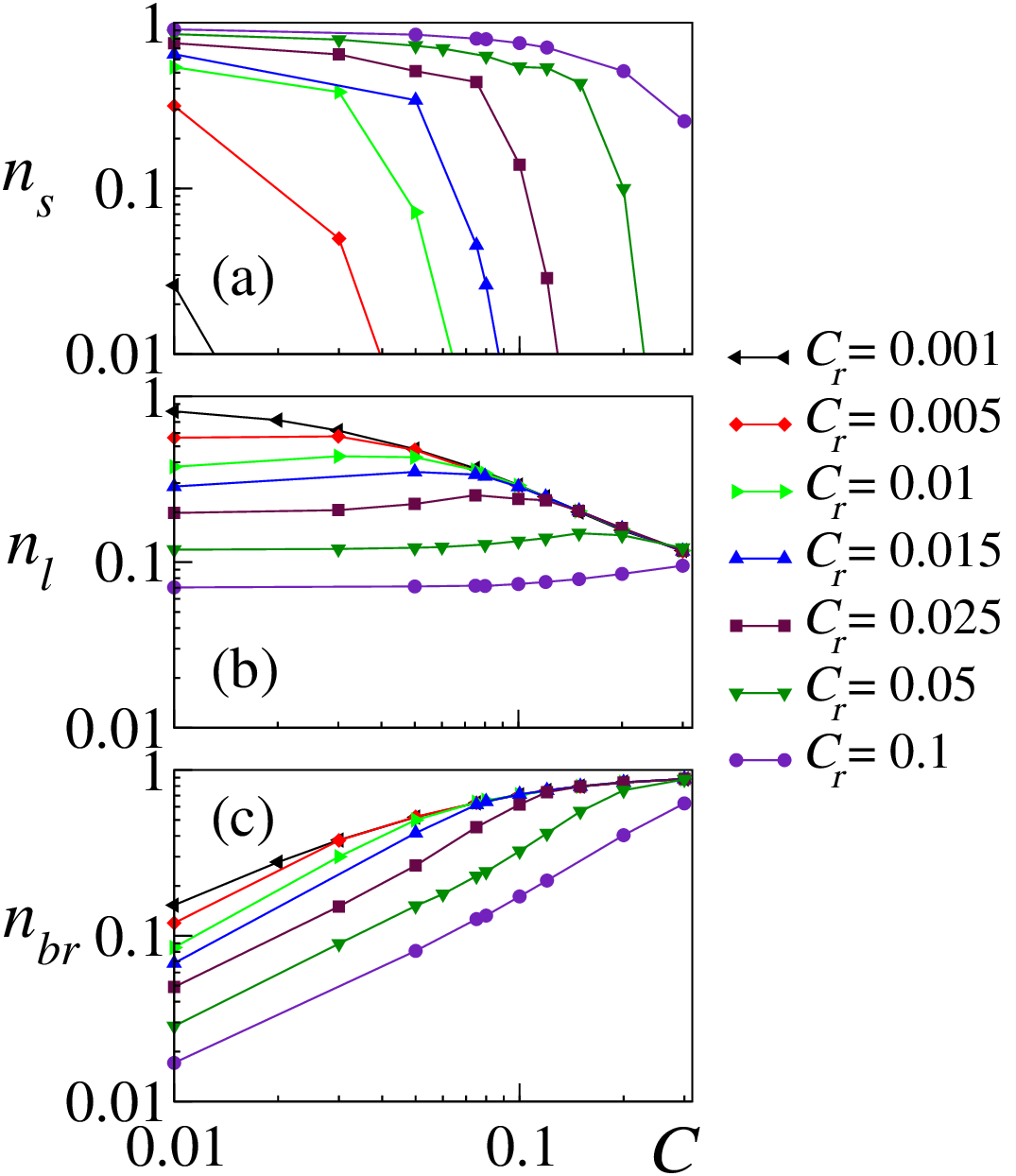}
\caption{Normalized number of single bonds reactant-chain $n_s$ (a),
loops $n_l$ (b) and bridges $n_{br}$ (c) as function of gelatin
concentration for $n_{as}=5$, $p_b=1$ and different values of
$C_r$.} \label{fig2}
\end{figure}

To better understand these results, we measure the number of
unsaturated reactants, the number of loops and bridges at the end of
the reaction. Such information, not accessible experimentally, is
extremely relevant to the mechanical properties of the gel. In
Fig.\ref{fig2} we plot the normalized number of single bonds $n_s$,
corresponding to the number of unsaturated reactants, the normalized
number of loops $n_l$ and the normalized number of bridges $n_{br}$
(defined at the end of section \ref{num}), as a function of the
chain concentration $C$ for different values of $C_r$.

We can detect three different regimes: i) At low $C_r$, when the
system is in the sol phase, the majority of bonds are loops at low
$C$ and bridges at high $C$; ii) in the intermediate regime the
number of loops decreases in favor of bridges leading to gelation;
iii) at high $C_r$ the majority of bonds are single bonds
crosslinker-chain, which brings the system back into the sol phase.

As a small remark, we notice that at very low $C_r$, corresponding
to the sol phase, $n_l$ decreases with $C$ over the whole observed
range following a power-law decay $n_l \sim C^{-0.8}$, in agreement
with previous results \cite{noi}. For increasing $C_r$, the same
power law decay is recovered at higher $C$. The analysis of $n_l$ in
systems with different pH indicates that a higher fraction of loops
is measured at a higher pH, as expected.
\begin{figure}[h]
\vspace{0.8cm}
\includegraphics[width=8cm]{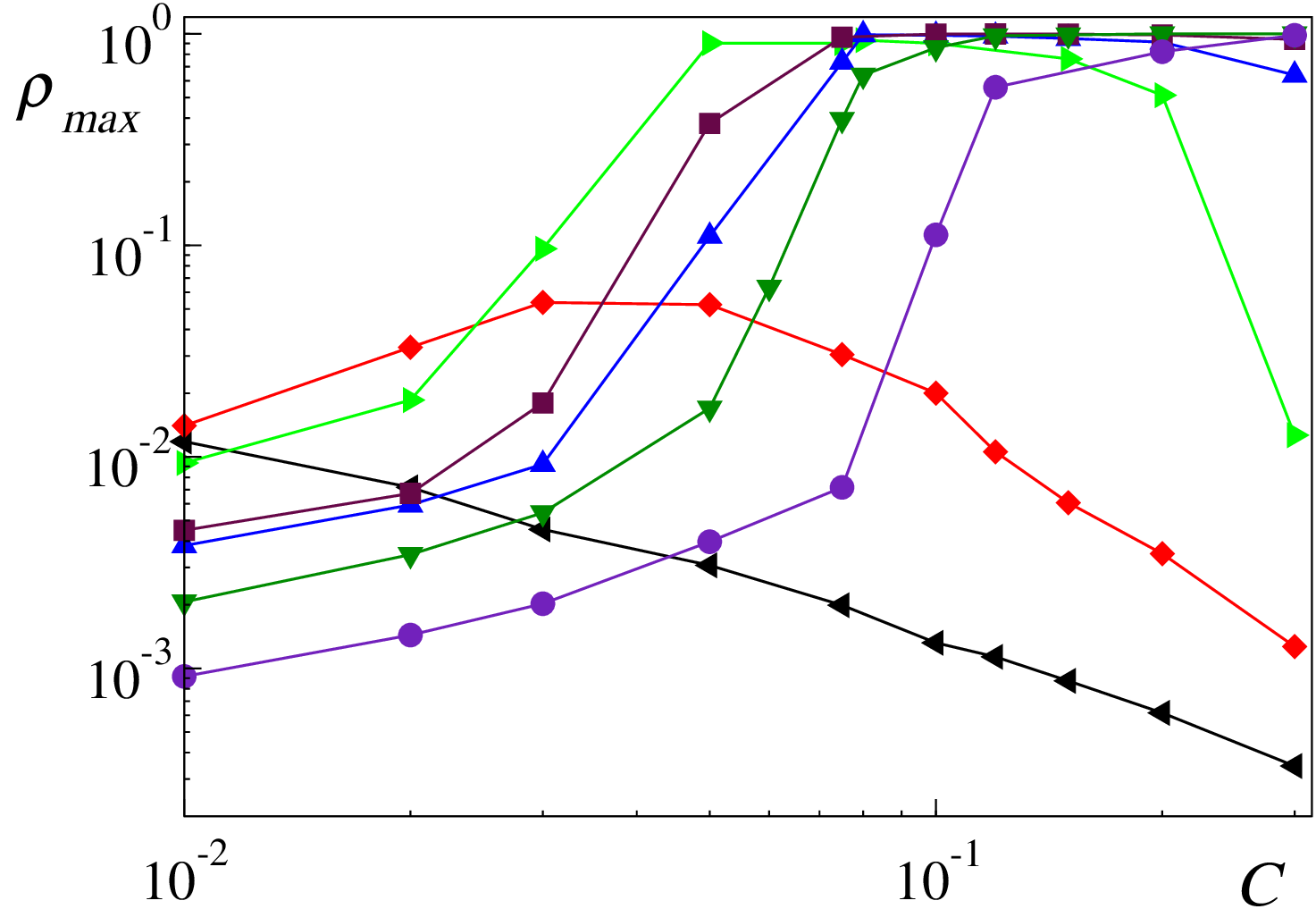}
\caption{Fraction of monomers in the largest cluster for $n_{as}=5$
and different reactant concentrations, ranging from $C_r=0.001$ to
$C_r=0.1$. The symbols indicate the same values of $C_r$ as in
Fig.\ref{fig2}.} \label{fig3}
\end{figure}

\begin{figure}
\includegraphics[width=7cm]{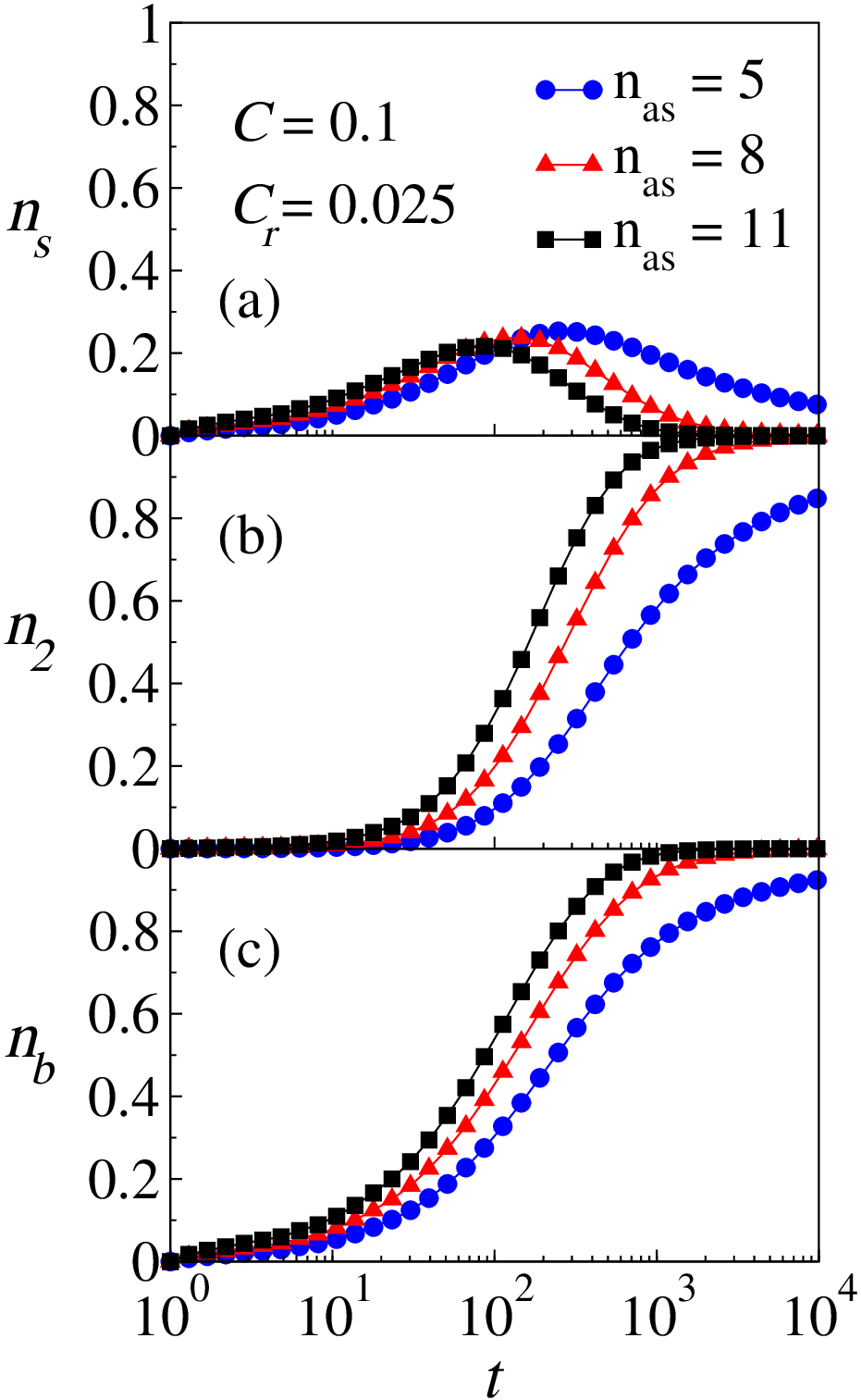}
\caption{(Number of single-bonds (a), number of bonds forming
bridges or loops (b) and total number of bonds (c), normalized by
the total number of possible bonds, as a function of time for
$C=0.1$, $C_r=0.025$ with $n_{as}=5$ (circles), $8$ (triangles),
$11$ (squares) and $p_b=1$. At the end of the reaction all systems
are in the gel phase.} \label{fig5}
\end{figure}

We also measure the fraction of monomers belonging to the largest
cluster, $\rho_{max}=s_{max}/N$, where $N=(C+C_r)L^3/8$ is the total
number of monomers in the system. Here $s_{max}$ is evaluated by
counting monomers, either in a chain or reactants. In Fig.\ref{fig3}
$\rho_{max}$ is plotted as a function of $C$. Except for the very
small $C_r$, corresponding to the sol phase, the fraction of
monomers belonging to the largest cluster increases toward unity for
increasing $C$, as the system moves into the gel phase. For higher
$C$, $\rho_{max}$ exhibits the tendency to decrease, as the system
goes back into the sol phase.

\emph{\textbf{Discussion:}} Depending on specific conditions, the
complex balance between $C$, $C_r$ and $n_{as}$, results in
enhancing the probability of forming either loops or single bonds
reactant-chain. For all pH values and low $C_r$ gelation is not
observed as reactants mostly form loops within the same chain.
Conversely, gelation is hindered at high $C_r$ since reactants
saturate all the active sites of the chains, not leading to the
formation of a macromolecule. Indeed this effect is observed at
higher $C_r$ for higher $n_{as}$, since, at higher pH, more
reactants are needed to saturate all active sites. Therefore, beyond
a certain value of $C_r$, adding reactants to the solution is
useless, as a higher concentration of chains is needed for gelation.
\\In the following section we investigate how the complex scenario
above described is connected to the gelation kinetics.

\section{Kinetics of bond formation}
\label{bond} In experiments the pH of the solution is found to
strongly influence not only the final state of the system, i.e. the
phase diagram, but also the kinetics of bond formation.

\begin{figure}
\includegraphics[width=7cm]{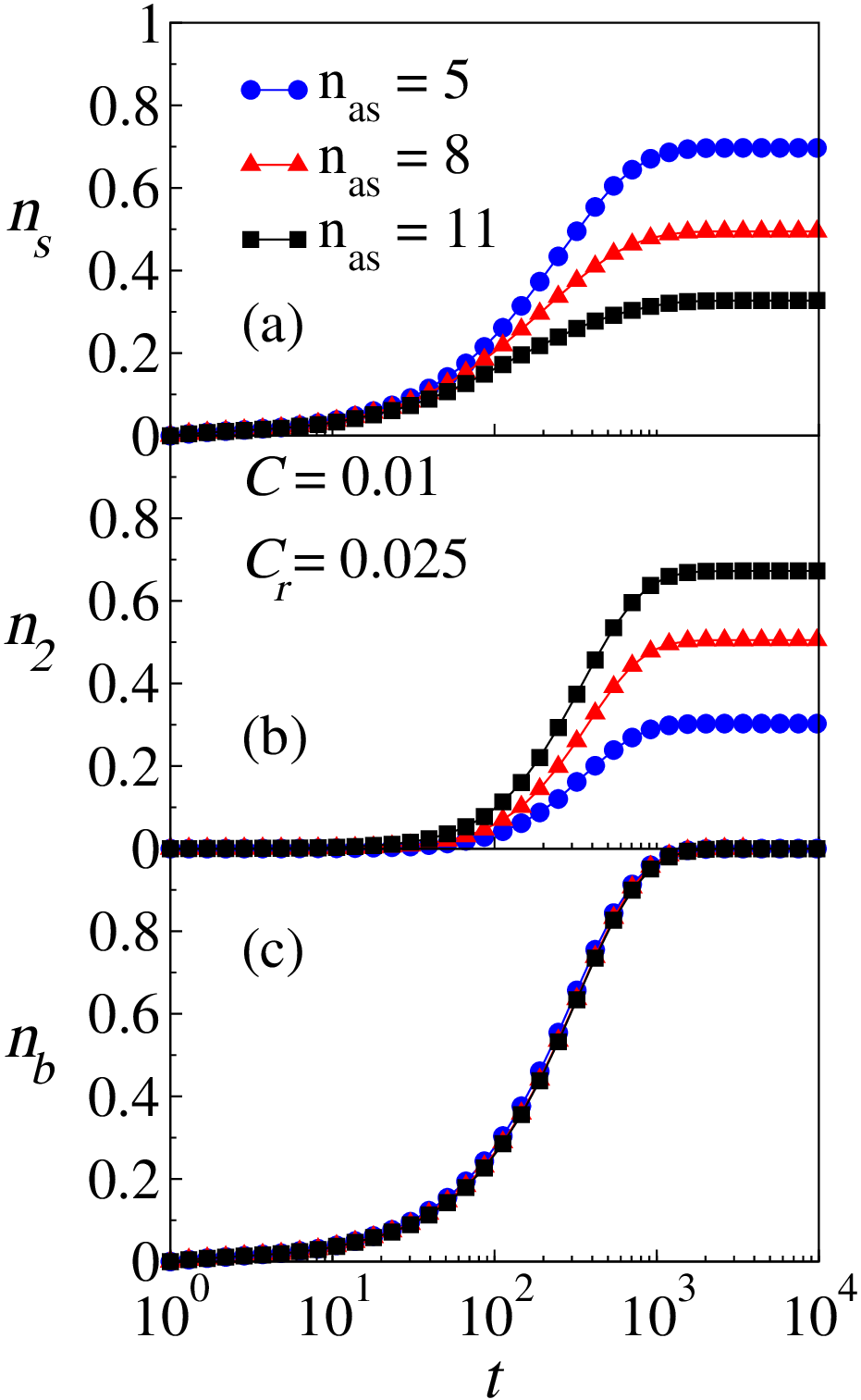}
\caption{Number of single-bonds (a), number of bonds forming bridges
or loops (b) and total number of bonds (c), normalized by the total
number of possible bonds, as a function of time for $C=0.01$,
$C_r=0.025$ with $n_{as}=5$ (circles), $8$ (triangles), $11$
(squares) and $p_b=1$. At the end of the reaction all systems are in
the sol phase. } \label{fig5_bis}
\end{figure}
In order to investigate this phenomenon, during the
reaction we monitor the total number of bonds formed $n_b(t)$, the number of
single-bonds $n_{s}(t)$ (bonds of type 1) and the number of bonds
forming bridges or loops $n_2(t)$ (i.e. bonds of type 2 and 3), with
obviously $n_b(t)=n_s(t)+n_2(t)$. The number of bonds is normalized
by the total number of possible bonds.

In Fig.\ref{fig5} $n_{s}(t)$, $n_2(t)$ and $n_b(t)$ are plotted as a
function of time in the case $p_b =1$ for $n_{as}=5,8,11$ for
$C=0.1$ and $C_r=0.025$, corresponding to the gel phase at the end
of reaction. As the reaction begins, single-bonds form rapidly, then
second bonds start to form and the degree of connectivity between
chains increases: single-bonds decrease in time as the number of
bridges increases. Our data indicate that as $n_{as}$ increases,
second bonds form faster, whereas the characteristic time of
single-bond formation is less affected by pH. As a consequence, we
expect that the ratio of the two times decreases with pH. A
significantly different behavior is observed in the sol phase
(Fig.\ref{fig5_bis}): in this case the characteristic time for the
first and the second bond formation are the same for all pH values.
Indeed, in this regime the whole kinetics of bond formation is
independent of the solution pH (see Fig.\ref{fig5_bis}c). Hence, for
$p_b=1$, the time ratio is expected to be almost constant by varying
the pH.

\begin{figure}
\includegraphics[width=7cm]{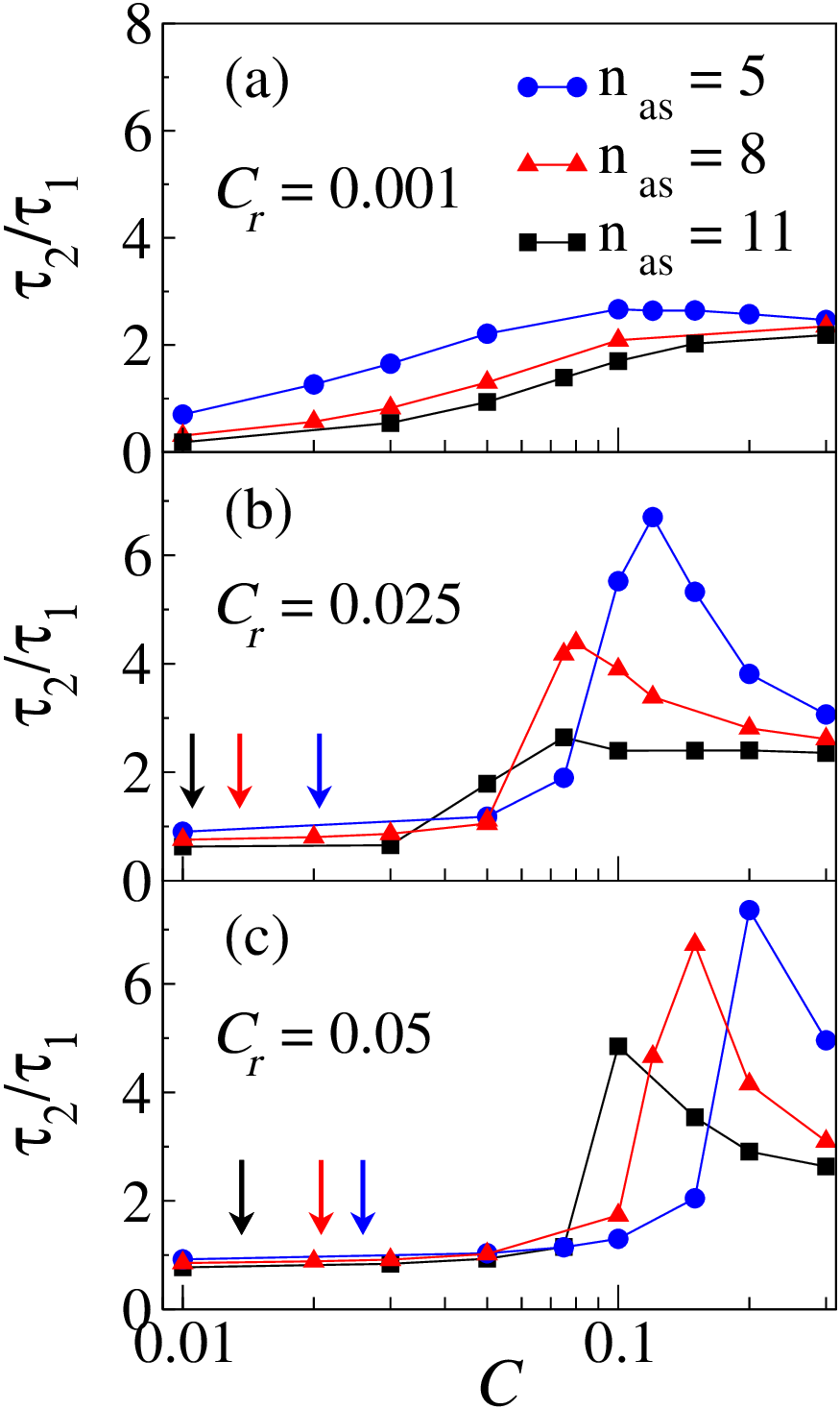}
\caption{Ratio between the two characteristic times for
$n_{as}=5,8,11$, $p_b=1$ and for $C_{r}=0.001$ (a) $C_r=0.05$ (b)
and $C_{r}=0.05$ (c). The arrows indicate the value of gelatin
concentration where gelation occurs for the different numbers of
active sites. } \label{fig6}
\end{figure}

We now focus on the ratio between the characteristic times $\tau_1$
of single bond formation and $\tau_2$ of second bond formation and
we study its dependence as a function of the number of active sites
per chain $n_{as}$. In Fig.\ref{fig6} we distinguish different
regimes, according to the concentration of chains and cross-linkers.
For low $C_r$ (Fig.\ref{fig6}(a)) all systems are in the sol phase
due to the very high number of loops (Fig.\ref{fig2}) and therefore
the time ratio is close to one. As the number of active sites per
chain increases, $\tau_1$ remains quite constant whereas $\tau_2$
decreases, so that $\tau_2/\tau_1$ slightly decreases as $n_{as}$
increases. For intermediate cross-linker concentrations
(Fig.\ref{fig6}(b)) and low chain concentrations (sol phase),
neither $\tau_1$ nor $\tau_2$ strongly vary increasing $n_{as}$,
consequently the ratio $\tau_2/\tau_1$ is almost independent of
$n_{as}$ and weakly decreases increasing $n_{as}$. For higher chain
concentrations, beyond the gelation threshold (indicated by an arrow
in the figure), the time ratio starts to increase, due to the slower
decrease of $\tau_2$ with respect to $\tau_1$, caused by the
presence of the macromolecule. The smaller the number of active
sites per chain, the later the increase of $\tau_2/\tau_1$ is
observed with the chain concentration, in agreement with the
displacement of the gel line in the phase diagram of Fig.\ref{fig1}.
Moreover, for smaller number of active sites the height of the peak
decreases, suggesting that $\tau_{2}$ decreases faster than
$\tau_1$. For high $C_r$ (Fig.\ref{fig6}(c)) the position of the
maximum time ratio appears to move toward higher chain
concentrations, due to the shifting of the gelation transition.

\begin{figure}
\includegraphics[width=7cm]{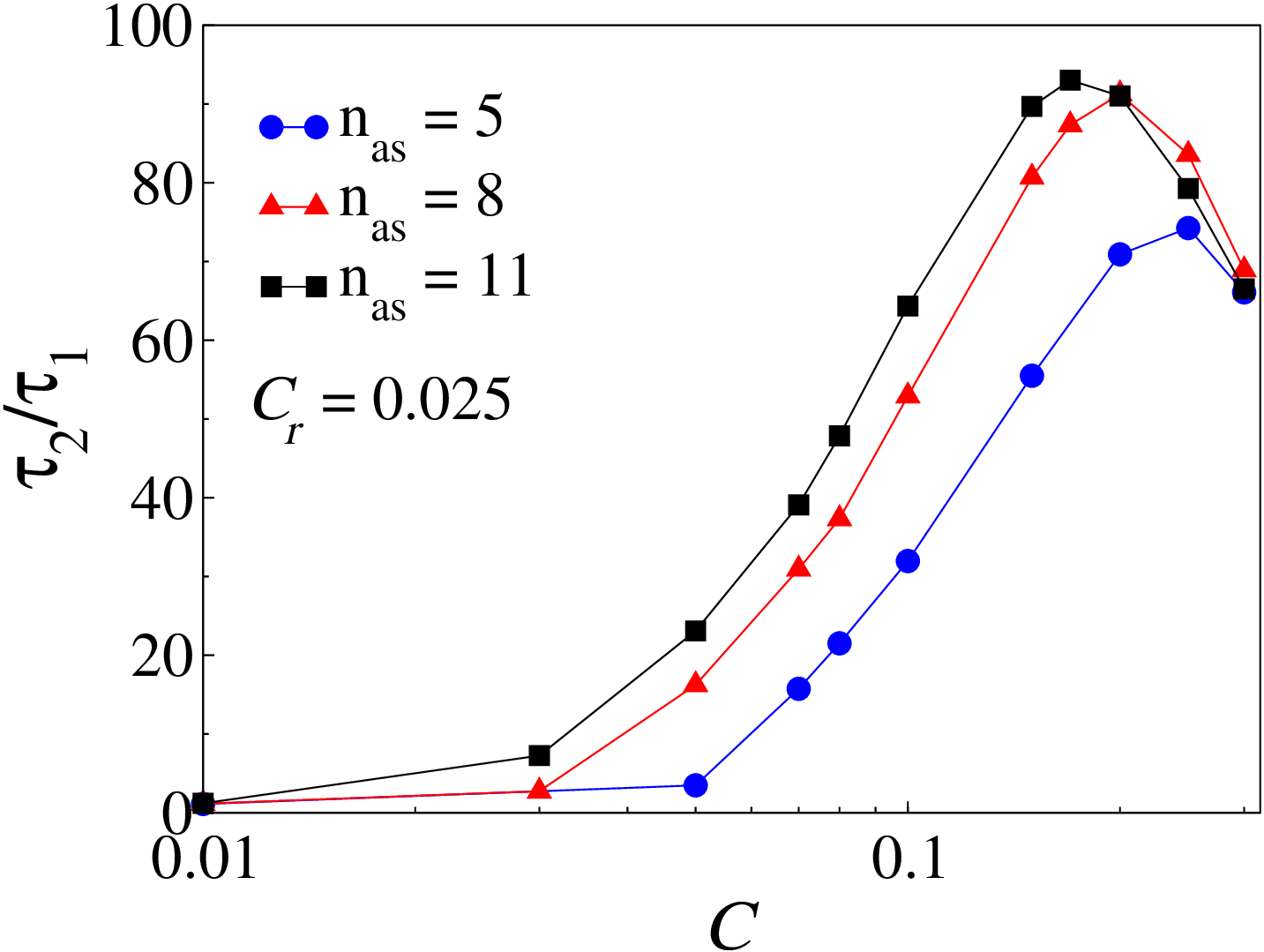}
\caption{ Ratio between the two characteristic times for $n_{as}=5$,
$C_{r}=0.025$ and $p_b=0.001$ as a function of chain concentration
$C$. } \label{fig7}
\end{figure}

We conclude our study by investigating the interplay between the
cross-linker reactivity and the pH. A first important hint of the
possible combined effects is obtained by comparing the data of
Fig.\ref{fig6}(b), discussed above and corresponding to $p_{b}=1$,
to the behavior of $\tau_2/\tau_1$ as a function of $C$ measured for
the same $C_r=0.025$ but for a lower reactivity reactant,
$p_b=0.001$ (Fig.\ref{fig7}). In this case the time ratio is about
one order of magnitude larger than for high reactivity crosslinkers
(Fig.\ref{fig6}). Moreover, the dependence of $\tau_2/\tau_1$ on the
number of active sites seems to be less strong and qualitatively
different. In fact, in contrast to what observed in
Fig.\ref{fig6}(b), here in the gel phase $\tau_2/\tau_1$ increases
with increasing $n_{as}$, since $\tau_1$ decreases whereas $\tau_2$
is very large for all $n_{as}$ due to the very small $p_b$.

\begin{figure}
\includegraphics[width=7cm]{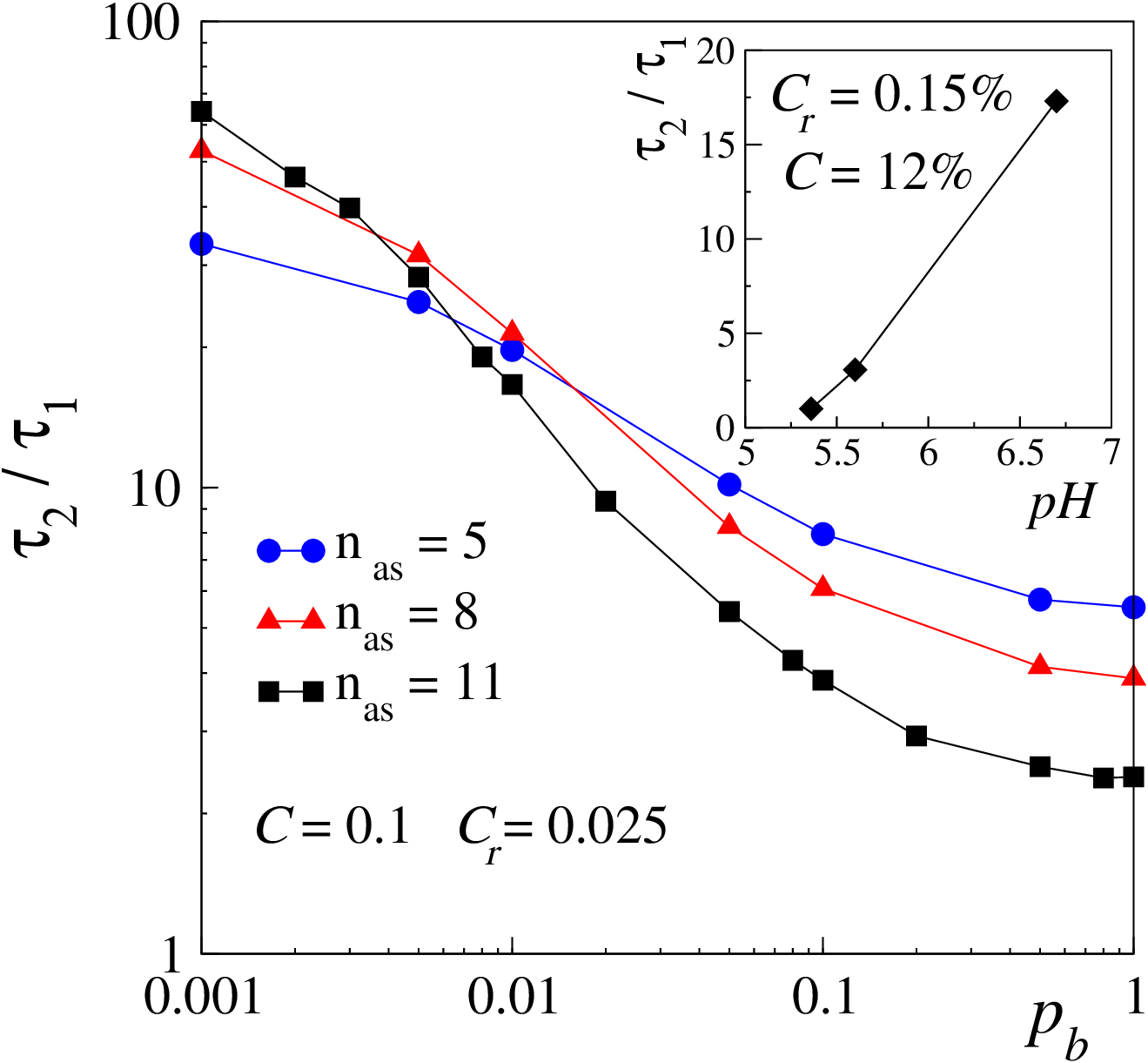}
\caption{Ratio between the two characteristic times for
$n_{as}=5,8,11$, $C_{r}=0.025$ and $C=0.1$ as a function of bond
probability $p_b$. For all $n_{as}$ the system is in the gel phase.
Inset: Ratio between the two characteristic times as a function of
solution pH in of gelatin chains in solution with BVSM.}
\label{fig8}
\end{figure}
In general, if the probability $p_b$ of forming the second bond
varies, the mean time of second bond formation $\tau_2$ changes, and
so does the velocity of the reaction. In Fig.\ref{fig8} we plot the
ratio $\tau_2/\tau_1$ as a function of $p_b$ for different values of
$n_{as}$. It is worth to notice that the average $\tau_1$ is
constant along each line, since $p_b$ only affects $\tau_2$. If the
bond probability is very low ($p_b \lesssim 0.005$) an increase of
$n_{as}$ will correspond to an increase of $\tau_2/\tau_1$, in
agreement with Fig.7. In this case the ratio is controlled by
$\tau_1$, which decreases as the number of active sites increases.
Conversely, if $p_b$ is sufficiently high ($p_b \gtrsim 0.01$) we
observe the opposite behavior (see Fig.6(b)). In this case, the
ratio is controlled by $\tau_2$: As the number of active sites per
chain increases, the mean time to form a second bond $\tau_2$
decreases more rapidly, affecting the ratio $\tau_2/\tau_1$.

{\bf \emph{Discussion:}} Our data indicate that when the system at
the end of reaction is in the sol phase, the ratio between the
characteristic times does not substantially vary with the pH.
Conversely, in the gel
phase the kinetics of bond formation is strongly influenced by the
$n_{as}$, i.e. by the pH. Since the time ratio $\tau_2/\tau_1$ is
actually measurable from the experiments \cite{noi}, our observation
suggests a new alternative way to discriminate the sol from the gel
phase resulting from crosslinking the gelatin solution in specific
conditions of $C$ and $C_{r}$, purely based on the reaction
kinetics.

It is worth to remark that the presence of loops, in general,
corresponds to smaller values of $\tau_2$, whereas the formation of
bridges leads to larger values, as follows from the microscopic
interpretation  of $\tau_1$ and $\tau_2$ \cite{noi}. At high $C$ the
effect of the pH in very dense systems becomes less important.
Indeed, at high $C$ both $\tau_1$ and $\tau_2$ decrease as $n_{as}$
increases, so that the ratio $\tau_2/\tau_1$ tends towards a value
which is independent of the number of active sites per chain. This
analysis allows to identify the range of parameters $C$ and $C_{r}$
for which the reaction kinetics is more sensitive to pH changes.
This is certainly an important information for the experimental
characterization of these systems. Moreover, it also suggests
possible new applications in terms of pH-responsive systems.

Furthermore, for low reactivity cross-linkers the kinetics of the
reaction is mainly controlled by the characteristic time for forming
the first bond, and therefore the time ratio increases with the pH.
For high reactivity cross-linkers, conversely, the combined effect
of the pH and of the reactivity makes $\tau_2$ decrease faster than
$\tau_1$ and therefore a decrease of the ratio will be observed when
$n_{as}$ increases (i.e. for higher pH). By measuring the dependence
of the time ratio on the pH, it is then possible to infer the level
of reactivity of the reactant. For instance, experimental data for
gelatin solutions with BVSM reactant \cite{mad3} show that, at fixed
$C$ and $C_r$, the time ratio increases with the pH (Inset
Fig.\ref{fig8}). According to our analysis this would suggest that
BSVM is a low reactivity cross-linker, namely that the number of
amine groups and reactant configurations leading to a chemical bond
is small. Interestingly enough, this finding does complete an
independent previous analysis \cite{noi}, which suggested that the
peculiar two-time kinetics observed in these systems should
correspond to a low value of the crosslinker reactivity $p_{b}$.

\section{Conclusions}
\label{conclu} In conclusion, we have performed a comprehensive
study of the pH dependence of the phase diagram and the reaction
kinetics in crosslinking gelatin solutions. This study is based on
numerical simulations of a lattice model specifically developed for
these systems. The results allow to rationalize the experimental
findings and give new relevant insights. The first important result
of our study is that the increase of the cross-linker concentration
does not necessarily imply the presence of a gel at the end of
reaction: indeed it may also hinder the gelation transition,
inducing the formation of a high number of single bonds or loops.
This effect moves towards higher concentrations if the pH of the
solution increases. As a consequence, we observe a reentrant phase
diagram.

Our study points out that also the kinetics of bond formation is
strongly influenced by the pH of the solution, with different
regimes according to the concentrations of chains and cross-linkers.
We find a remarkable qualitative difference in the pH dependence of
the reaction kinetics, which could be used as a new purely kinetic
criterion to discriminate between the sol and the gel final state of
the crosslinking process. Thanks to a detailed analysis of reaction
kinetics in terms of single bonds, loops within chains and bridges
between different gelatin chains, we are able to give a meaningful
explanation of the different regimes observed. Most interestingly,
our results show that there exist ranges of the control parameters
$C$ and $C_{r}$ for which the behavior of the system may be
extremely sensitive to pH changes. These findings are fundamentally
important for the experimental studies and suggest new possible
applications of these materials.

Finally, we have been able to clarify how the influence of the pH on the
kinetics of bond
formation also depends on the reactivity of cross-linker considered, i.e.,
on the bond probability $p_b$. For low reactivity cross-linkers the kinetics of
the reaction is mainly controlled by the single-bond formation time,
which decreases with pH, leading to an increase in the time ratio.
Conversely, for high reactivity cross-linkers and for high
chain concentration, the increase of pH induces
a definite decreases in the time ratio.

\end{document}